\documentclass[two column]{aastex631}

\usepackage{subfigure}
\usepackage{amsmath}  
\usepackage{graphicx}
\usepackage{array}
\usepackage{hyperref}
\usepackage{upgreek}
\defcitealias{Sawala2025}{S25}
\shorttitle{LG fate}
\shortauthors{Wu et al.}

\begin{document}
\title{The Fate of the Milky Way–Andromeda System: To Merge or Not?} 
\author[0009-0006-7193-4443]{Hao Wu}
\affiliation{Department of Astronomy, School of Physics, Peking University, Beijing 100871, China}
\affiliation{Kavli Institute for Astronomy and Astrophysics, Peking University, Beijing 100871, China}

\author[0000-0003-3250-2876]{Yang Huang}
\affiliation{School of Astronomy and Space Science, University of Chinese Academy of Science, Beijing 100049, China}
\affiliation{National Astronomical Observatories, Chinese Academy of Sciences, Beijing 100101, China}
\author[0000-0002-7727-1699]{Huawei Zhang}
\affiliation{Department of Astronomy, School of Physics, Peking University, Beijing 100871, China}
\affiliation{Kavli Institute for Astronomy and Astrophysics, Peking University, Beijing 100871, China}
\author{Guangze Sun}
\affiliation{Department of Astronomy, School of Physics, Peking University, Beijing 100871, China}
\affiliation{Kavli Institute for Astronomy and Astrophysics, Peking University, Beijing 100871, China}
\author{Shi Shao}
\affiliation{National Astronomical Observatories, Chinese Academy of Sciences, Beijing 100101, China}
\received{29-Jan-2026}
\revised{17-Mar-2026}
\accepted{24-Mar-2026}
\correspondingauthor{Huawei Zhang: zhanghw@pku.edu.cn;\\Yang Huang: huangyang@ucas.ac.cn}
\begin{abstract}
It has long been predicted that the Milky Way (MW) will eventually merge with Andromeda (M31), a view reinforced by \textit{HST} measurements indicating a small M31 transverse velocity. However, using updated \textit{Gaia}-based proper motions (PMs) and including the dynamical influence of the Large Magellanic Cloud (LMC) and M33, Sawala et al. reported an MW–M31 merger probability of $\sim$50\% within 10 Gyr, leaving the fate of the Local Group (LG) uncertain. Adopting their semi-analytic framework, we revisit this problem with the latest and most precise \textit{Gaia}-based PMs for M31 and M33, corrected for systematic offsets in \textit{Gaia} astrometry. In our fiducial model, the MW–M31 merger probability rises to 90\%, with a median merger time of $6.5_{-1.5}^{+1.3}$ Gyr, broadly restoring the classical picture. A sensitivity analysis shows that the merger probability depends strongly on the adopted M31 PM through two channels: a direct effect via the radial-tangential balance of the MW-M31 orbit, and a satellite-mediated effect, where the M31 PM fixes the orbital plane and determines how satellite-induced barycentric reflex motions project onto it, either promoting or suppressing a merger. Given this sensitivity, current measurements, while favoring a high merger probability, remain inconclusive, spanning from 64.7\% to 100\% across the 2$\sigma$ PM region. Future PM measurements with uncertainty of $\lesssim2\,\upmu\mathrm{as\,yr^{-1}}$ will be required to reach a firm conclusion, i.e., to constrain the probability range within 10\% at the 2$\sigma$ level.
\end{abstract}

\section{Introduction}\label{introduction}
The Milky Way (MW) and Andromeda (M31), together with their satellite systems, constitute the Local Group (LG). The LG provides a unique laboratory for investigating hierarchical galaxy formation and evolution \citep[e.g.,][]{Hammer2018,Helmi2020} and also serves as an important benchmark for cosmological models. Two remarkable properties of the M31 satellite system, the presence of a vast, thin, co-rotating plane and the significant lopsidedness of the satellite distribution, pose challenges to the standard cosmological paradigm \citep[e.g.,][]{Ibata2014,Kanehisa2025}.

Dynamical evolution of the most massive LG members, the MW and M31, and their most massive satellites, the Large Magellanic Cloud (LMC) and the Triangulum Galaxy (M33), is fundamental to understanding the LG itself, from constraining its total mass to explaining the origin of tidal substructures within its dominant galaxies \citep[e.g.,][]{vdm2012a,McConnachie2009}. Given M31's large heliocentric radial velocity of around $-300$ km\,s$^{-1}$ (corresponding to $\sim-110$ km\,s$^{-1}$ in the Galactocentric frame) known for over a century \citep{Slipher1913}, the scenario of an eventual MW–M31 collision has been explored even without precise M31 proper motion (PM) measurement \citep{Dubinski1996,Cox2008}. 
The first direct proper-motion (PM) measurement of M31 was obtained with the Hubble Space Telescope (HST) by \citet{vdm2012a}. By combining this measurement with earlier indirect constraints from the motions of M31 satellites \citep{vdm2008}, they derived a transverse velocity of $17 \pm 17\,\mathrm{km\,s^{-1}}$ for M31 relative to the MW \citep{vdm2012a}.
Based on this, together with the precisely measured M33 PM from Very Long Baseline Array (VLBA) water-maser observations \citep{Brunthaler2005}, \citet{vdm2012b} provided a systematic analysis of the future evolution of the MW–M31–M33 system through $N$-body simulations and semi-analytic integrations. They found that a MW–M31 merger is essentially inevitable, occurring in $5.86^{+1.61}_{-0.72}$ Gyr from now. These results have since established the MW–M31 collision scenario as the widely accepted picture for the fate of the LG. 

\begin{figure}[htbp!]
  \centering
    \includegraphics[width=0.9\linewidth]{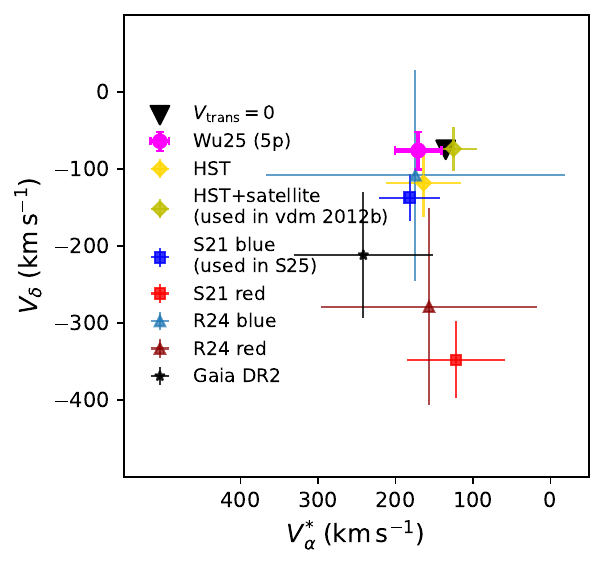}
  \caption{Comparison of literature measurements of the M31 transverse velocity. The black inverted triangle indicates a purely radial orbit of M31 relative to the MW. The magenta symbol shows the result of \citet{Wu2025} based on {\it Gaia} stars with five-parameter (5p) astrometric solutions. The yellow symbol denotes the HST measurement from \citet{vdm2012a}, while the olive symbol shows the value obtained after combining it with the earlier satellite-based estimate of \citet{vdm2008}. 
  Blue and red squares show the blue- and red-subsample results of \citet{Salomon2021}, with the blue value adopted by \citetalias{Sawala2025}. Light-blue and dark-red triangles indicate the corresponding results of \citet{Rusterucci2024}. The black star marks the earlier {\it Gaia} DR2 estimate from \citet{vdm2019}.}
  \label{fig:m31_pm_compare}
\end{figure}

Recently, however, two independent developments have made the inevitability of a MW-M31 collision less definitive. The first is the discrepancy between the {\it Gaia}-based PMs of M31 and M33 and those obtained with HST/VLBA. A detailed comparison is shown in Fig.~\ref{fig:m31_pm_compare}. \citet{vdm2019} demonstrated the feasibility of measuring the PMs of M31 and M33 using astrometry from {\it Gaia} Data Release 2 ({\it Gaia} DR2; \citealt{Gaiadr2}), but their inferred transverse velocities differ significantly from the HST/VLBA values. Using the improved astrometry from {\it Gaia} Early Data Release 3 ({\it Gaia} EDR3; \citealt{Gaiaedr3}), \citet{Salomon2021} refined the M31 PMs and found that the results are highly sensitive to the adopted stellar samples: PMs derived from the blue and red subsamples differ by more than 200\,km\,s$^{-1}$. This sample dependence was further examined by \citet{Rusterucci2024}, which constructed a quasar-based reference frame and re-analysed the same stars in four spatial regions. They confirmed that discrepancy between the blue and red samples persists and cannot be explained by spatial selection effects. 
Even the blue-sample result in \citet{Salomon2021}, which best matches the HST value, yields a transverse velocity exceeding 80\,km\,s$^{-1}$, much larger than the value used in \citet{vdm2012b}, implying a less strongly radial MW-M31 relative motion.
The second development concerns the LMC. A first-infall scenario, in which the LMC experienced a recent pericentric passage at about 50\,kpc, is now widely favored \citep[e.g.,][]{Besla2007,Kallivayalil2013,Patel2017,Patel2020}. Such a massive ($\sim1.5\times10^{11}\,M_{\odot}$; \citealt{Erkal2021}) recent infall naturally induces a reflex motion of the MW \citep[e.g.,][]{Conroy2021,Camargo2021,Sheng2024}, displacing the barycenter of the MW-M31 system and modifying predictions for their future encounter. \citet[][hereafter \citetalias{Sawala2025}]{Sawala2025} combined newly derived {\it Gaia} PMs of M31 and M33 and included the LMC when modeling the fate of the LG. Using semi-analytic orbit integrations with observational uncertainties sampled to $2\sigma$, they found that the MW-M31 merger probability was reduced by roughly half, with the LMC-induced reflex motion of the MW increasing the velocity component perpendicular to the MW-M31 orbital plane, thereby suppressing the merger probability.

PMs are essential for predicting the future evolution of the LG. However, the discrepancy between M31 PM measurements derived from the {\it Gaia} blue and red stellar samples has introduced substantial uncertainty into such predictions. In our previous work, \citet{Wu2025} demonstrated that this discrepancy is caused by systematic offsets between the {\it Gaia} five-parameter (5p) and six-parameter (6p) astrometric solutions. After correcting for these systematics, the PM measurements from the blue and red samples agree within $1\sigma$ \citep{Wu2025}. This correction enabled precise PM measurements for both M31 and M33 based on stars with high-quality 5p astrometry. The derived M31 transverse velocity, $46.7\pm24.0\,\mathrm{km\,s^{-1}}$, is smaller than the values reported by \citet{Salomon2021} and adopted in \citetalias{Sawala2025}. Building on these improved measurements, we reassess the future dynamical evolution of the LG under the updated PM constraints and quantify how the revised M31 and M33 PMs influence the outcome of the MW-M31 encounter. Section~\ref{sec:method} describes the methodology and initial conditions for the semi-analytic integrations, Section~\ref{sec:result} presents the results and their implications, and Section~\ref{sec:summary} summarizes the main conclusions.

\section{Method}\label{sec:method}
We employ a semi-analytic framework to model the future dynamical evolution of the LG and to quantify the probability of an MW-M31 merger. Our approach largely follows \citetalias{Sawala2025}: observational uncertainties are sampled via Monte Carlo (MC) realizations and propagated through numerical orbit integrations, resulting in statistical distributions of possible outcomes for the LG’s fate. The setup is briefly summarized below, while full methodological details can be found in \citetalias{Sawala2025}.

\begin{deluxetable*}{lcccccc}
\tablecaption{Prior settings for the 20 parameters in the fiducial model. \label{tab:prior}}
\tablehead{
\colhead{Galaxy} &
\colhead{$M_{200}$ ($10^{11}\,M_{\odot}$)} &
\colhead{$c$} &
\colhead{$\mu$} &
\colhead{$\mu_{\alpha*}$ ($\upmu$as yr$^{-1}$)} &
\colhead{$\mu_{\delta}$ ($\upmu$as yr$^{-1}$)} &
\colhead{$V_{\rm LOS}$ (km s$^{-1}$)}
}
\startdata
MW   & $10.0\pm2.0$   & $10.0\pm2.0$   & ---            & ---        & ---        & --- \\
M31  & $13.0\pm4.0$   & $15.0\pm2.5$ & $24.407\pm0.032$ & $45.9\pm8.1$ & $-20.5\pm6.6$& $-301.0\pm1.0$ \\
M33  & $3.0\pm1.0$    & $10.0\pm2.0$   & $24.670\pm0.070$   & $30.6\pm5.7$ & $19.0\pm5.7$ & $-179.2\pm1.8$ \\
LMC  & $1.5\pm0.5$& $10.0\pm2.0$   & $18.477\pm0.026$ & $1910.0\pm20.0$  & $229.0\pm47.0$   & $262.2\pm3.4$ \\
\enddata
\tablecomments{Values of distance moduli and velocities are all given in the heliocentric frame. 
The halo masses ($M_{200}$) follow the values adopted by \citetalias{Sawala2025}, where the reasons for these choices are discussed. The concentration parameters $c$ are also taken from \citetalias{Sawala2025}, except for $c_{\rm M31}$, which is adjusted to ensure consistency with observational constraints.}
The distance moduli ($\mu$) and line-of-sight velocities ($V_{\rm LOS}$) are taken from \citetalias{Sawala2025}, based on values compiled from \citet{Li2021, Ou2023, Pietrzyski2019, Watkins2013, McConnachie2012}. Proper motions are updated to the most precise measurements available: for M31 we adopt the values from \citet{Wu2025}; for M33 we use the weighted average of \citet{Wu2025} and \citet{Brunthaler2005}; and for the LMC we adopt \citet{Kallivayalil2013}.
\end{deluxetable*}

The framework begins with MC sampling of 20 observables for a four-body system comprising the MW, M31, M33, and the LMC, with each galaxy modeled as a spherical Navarro–Frenk–White (NFW) halo. We neglect the Small Magellanic Cloud (SMC) because its low mass renders its impact on the LG fate negligible \citepalias{Sawala2025}. The sampled quantities are four halo masses ($M_{200}$), four concentrations ($c$), three distance moduli ($\mu$), three two-component PMs ($\mu_{\alpha}^{*}$ and $\mu_{\delta}$), and three line-of-sight velocities ($V_{\rm LOS}$). Parameter priors in our fiducial model largely follow \citetalias{Sawala2025}, with several updates to better reflect current observational constraints. First, we update the PMs of M31 and M33 to the most precise determinations available: for M31 we adopt the \citet{Wu2025} measurement based on {\it Gaia} sources with high-quality 5p astrometric solutions, and for M33 we use the weighted average of the {\it Gaia}-based value from \citet{Wu2025} (the 5p solution) and the VLBA water maser result from \citet{Brunthaler2005}. In addition, although \citetalias{Sawala2025} reported little sensitivity of the MW–M31 merger rate to the M31 concentration $c_{\mathrm{M31}}$, we adopt $c_{\mathrm{M31}}=15\pm2.5$ rather than $10\pm2$ used in \citetalias{Sawala2025} to better match observational constraints on M31 rotation curve (e.g., $c_{\mathrm{M31}}=20.1\pm2.0$ in \citealt{Chemin2009}; $c_{\mathrm{M31}}=12.0$ in \citealt{Corbelli2010}; $c_{\mathrm{M31}}=12.5$ in \citealt{Tamm2012}). 
The fiducial halo masses adopted from \citetalias{Sawala2025} correspond to a relatively low-mass MW-M31 configuration compared with several previous semi-analytic studies \citep[e.g.,][]{vdm2019,Patel2017,Patel2025}. We therefore explore a broader range of halo masses in Section~\ref{subsec:PM_sensi}.
Detailed priors for all 20 parameters are listed in Table~\ref{tab:prior}. In our fiducial model, each parameter is drawn from a Gaussian prior informed by current measurements and truncated at $\pm2\sigma$ level, similar to \citetalias{Sawala2025}. To transform each galaxy's heliocentric positions and velocities into the system's center-of-mass (COM) frame, we adopt a Galactocentric distance of the Sun of $R_0 = 8.122$ kpc \citep{GRAVITY2018} and a solar velocity relative to the Galactic center of (12.9, 245.6, 7.78) km s$^{-1}$ \citep{Reid2004,GRAVITY2018,Drimmel2018}, following the same convention as \citetalias{Sawala2025}. For each run, 10,000 MC samples are generated to ensure statistical robustness.

\begin{figure*}[htbp!]
  \centering
    \includegraphics[width=0.75\textwidth]{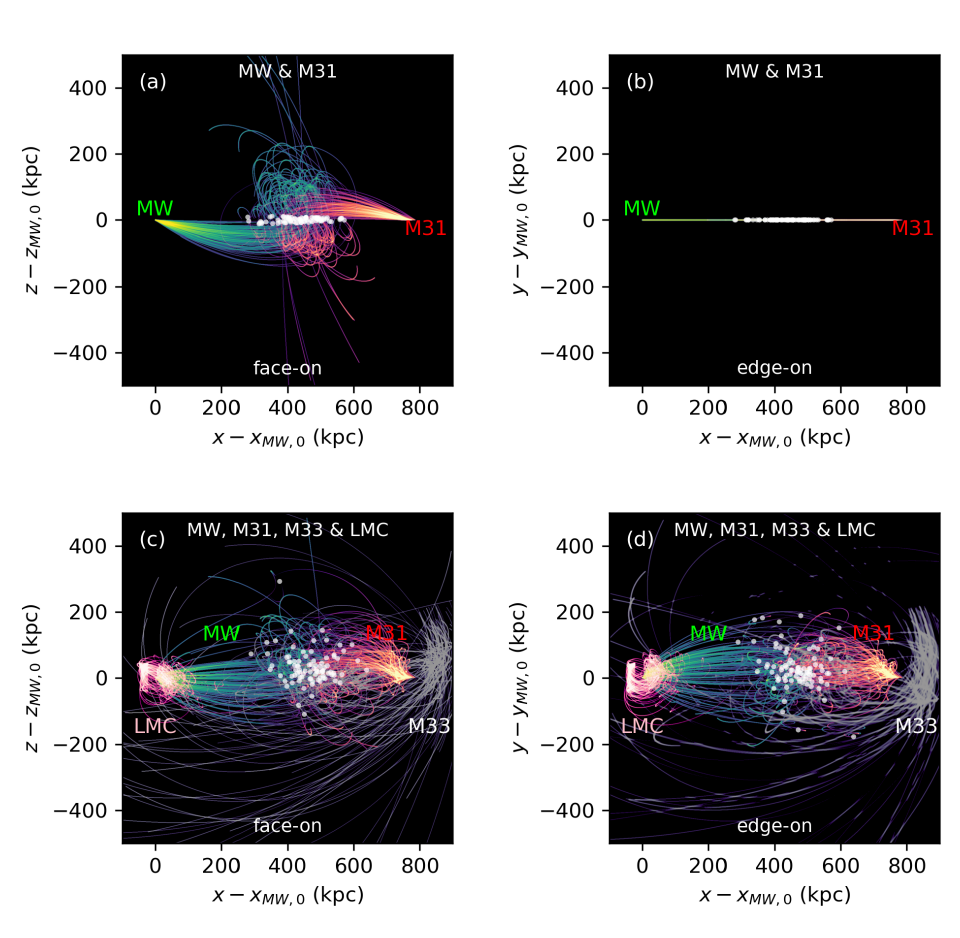}
  \caption{Predicted orbital evolution of the local group (LG) galaxies. Panels (a) and (b) show 100 randomly selected tracks from 10,000 MC realizations for the two-body configuration (MW-M31), and panels (c) and (d) show the same for the four-body configuration (MW, M31, M33 and LMC). In each case, the left panel is the face-on projection and the right panel is the edge-on projection with respect to the MW–M31 orbital plane, defined by the initial positions and velocities of the MW and M31, following \citetalias{Sawala2025}. The green and red lines trace the orbits of the MW and M31, while the pink and silver lines trace the LMC and M33, respectively. Line brightness indicates the relative probability density at any position. For each panel, the track of the satellites will terminate if a merger between them and the hosts is flagged, additionally, if a MW–M31 merger is flagged, a white point marks the merger location and the corresponding track terminates; otherwise, tracks are integrated to 10 Gyr.}
  \label{fig:4body_track}
\end{figure*}

With the sampled initial conditions, we integrate the orbits using the public code of \citetalias{Sawala2025} with its default settings \footnote{\url{https://github.com/TillSawala/MW-M31}}. The orbits are evolved in the COM frame with a leapfrog scheme and a 1 Myr timestep. 
Following \citetalias{Sawala2025}, the gravitational force between haloes is softened with a fixed length of $b=20$\,kpc to avoid unphysical hard scattering during close encounters and prevent an artificial suppression of the gravitational attraction. Tests in \citetalias{Sawala2025} showed that varying $b$ from 10 to 30\,kpc has little effect on the inferred merger statistics. We therefore adopt the same fiducial softening length for all four galaxies as in \citetalias{Sawala2025}.
Dynamical friction is implemented following the modified Chandrasekhar prescription of \citetalias{Sawala2025}; for comparable-mass pairs, the friction is distributed between the two bodies in proportion to their masses. A merger is flagged when the three-dimensional (3D) separation of a pair falls below 20 kpc, at which point the pair is replaced by a single remnant located at their common COM, and the system continues to evolve up to 10 Gyr.

In this work, our fiducial model integrates a four-body system (MW, M31, M33, LMC) across 10,000 MC realizations, with all parameters drawn from the Gaussian priors listed in Table~1 (truncated at $\pm2\sigma$). To isolate the influence of satellites on the MW–M31 merger probability, we additionally perform simulations for three-body configurations (MW, M31, M33 or LMC) and the two-body case (MW–M31), using identical sampling and integration settings. 
In the sensitivity tests (Section~\ref{subsec:PM_sensi}), only the priors of the parameters under investigation are modified, while all other priors and the integration procedure remain unchanged. For the PM tests, the priors of the M31 and M33 PMs are replaced with uniform distributions. For the mass tests, only the halo-mass priors being tested are varied.

\section{Results and Discussion}\label{sec:result}
\subsection{Orbital Evolution of the Local Group}\label{subsec:fiducial}
\begin{figure*}[htbp!]
  \centering
    \includegraphics[width=\linewidth]{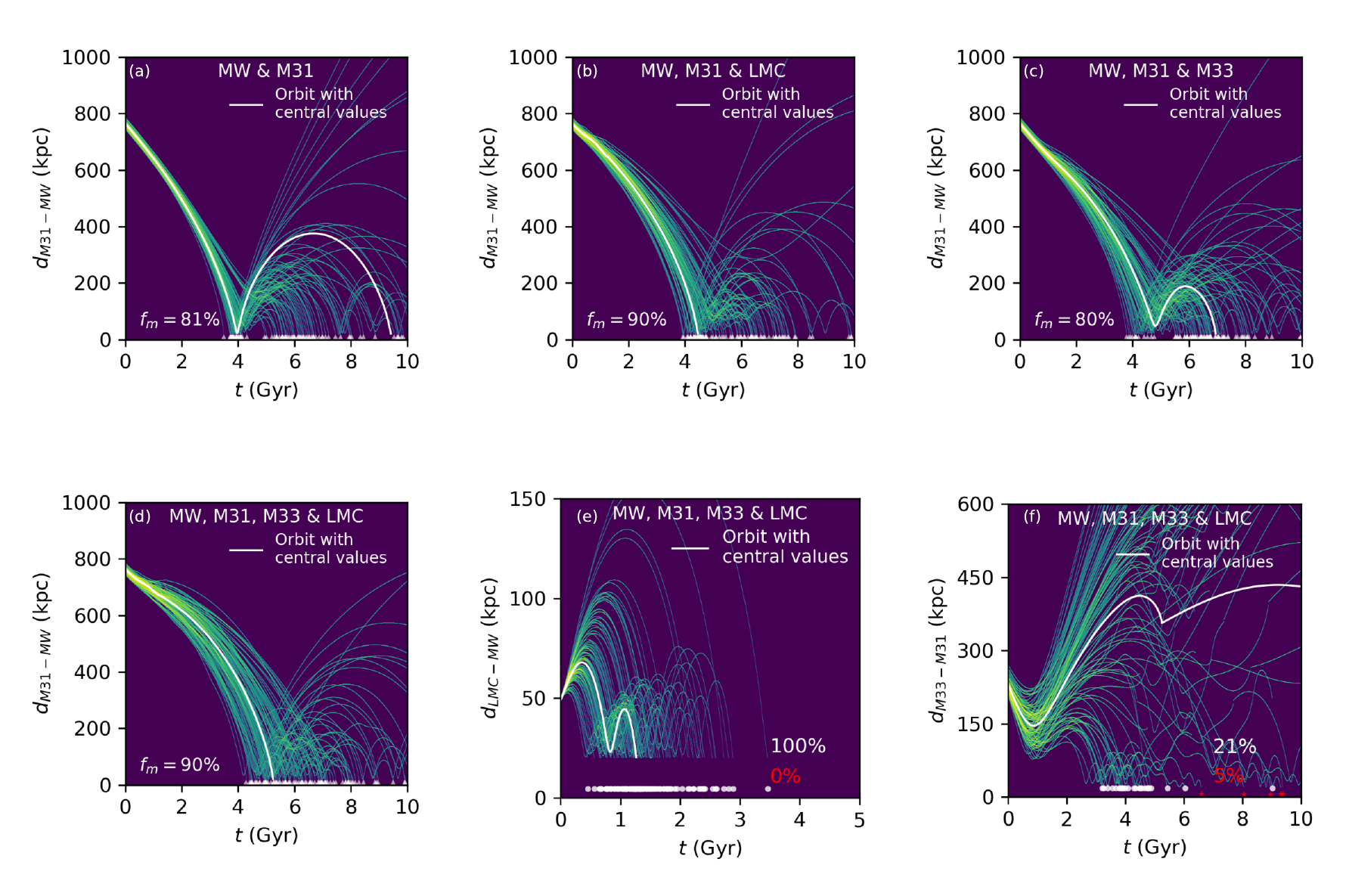}
  \caption{Panels (a)--(d): Time evolution of the 3D separation between the MW and M31 for two-, three-, and four-body configurations. Each panel shows 100 representative tracks randomly drawn from the 10{,}000 MC realizations. The MW–M31 merger fraction within 10 Gyr ($f_m$), computed from all 10{,}000 samples, is indicated in the lower-left corner. White symbols along the bottom mark the corresponding merger times, and the white curves show the evolution integrated using the central values of the fiducial priors for all 20 observables listed in Table~\ref{tab:prior}. Panels (e) and (f): Similar to panels (a)--(d), but showing the LMC–MW and M33–M31 separations in the fiducial four-body model. The merger fractions of each satellite with its host, before and after the MW–M31 merger, are indicated in white and red text, respectively. The corresponding merger times are marked by white and red symbols along the bottom axis of each panel.}
  \label{fig:distance_compare}
\end{figure*}

Starting from the MC-sampled initial conditions, we numerically integrate each realization following the procedure described in Section~\ref{sec:method}. Fig.~\ref{fig:4body_track} illustrates the orbital evolution of 100 randomly selected realizations from the full ensemble of 10,000 integrations for both the two-body and four-body configurations. Under the updated PM constraints, the MW–M31 relative motion becomes noticeably more radial than in \citetalias{Sawala2025}. The inclusion of M33 and the LMC introduces additional perturbations while further promoting the radial nature of the MW–M31 orbit.
Panels~(a)--(d) of Fig.~\ref{fig:distance_compare} show the time evolution of the 3D separation between M31 and the MW for 100 randomly selected realizations under different model setups: the two-body case (MW and M31), the three-body cases (MW, M31, and M33/LMC), and the fiducial four-body case (MW, M31, M33, and LMC). For the four-body configuration, panels~(e) and~(f) present the corresponding separations between LMC and MW, and between M33 and M31, respectively. Merger rates are derived from the complete set of 10,000 realizations.

These results provide a comprehensive picture of the future evolution of the MW–M31 system. In the baseline two-body configuration, the updated M31 PM implies a more radially dominated MW–M31 orbit than found by \citetalias{Sawala2025}, as illustrated in Fig.~\ref{fig:4body_track}, leading to a higher merger probability. When the full four-body system is considered, the MW–M31 merger fraction rises to 90\% among the 10,000 realizations, even higher than in the two-body case. This indicates that the two most massive members of the LG are very likely to merge within the next 10~Gyr. Our result therefore supports the earlier picture that the MW and M31 are expected to merge in the future \citep[e.g.,][]{vdm2012b}, rather than the $\sim$50\% merger probability reported by \citetalias{Sawala2025}.

Beyond the difference in merger probability, the roles of the satellites also change under the updated PMs of M31 and M33. In \citetalias{Sawala2025}, the MW’s reflex motion induced by the LMC adds a velocity component perpendicular to the MW–M31 orbital plane, thereby reducing the merger probability, while the interaction between M33 and M31 tends to increase it. With the updated PMs adopted in this work, however, the effect of the LMC is reversed relative to the result of \citetalias{Sawala2025}. Including the LMC, whose interaction and merger with the MW occur at a median time of $1.3_{-0.4}^{+0.6}$~Gyr from now and always precede the first MW–M31 encounter, drives the MW–M31 orbit to become even more radial, further increasing the merger probability relative to the two-body case. This behavior is primarily caused by the reflex motion induced by the massive LMC, which displaces the MW barycenter and enhances the radial component of the MW's velocity relative to M31.

By contrast, M33 exerts only a minor influence on the MW–M31 merger probability. Previous studies have examined the orbital history of M33 in detail and suggest that it is very likely on its first infall toward M31 \citep{Patel2017,Patel2025}. Such an infall may have induced a reflex motion of M31 over the past few Gyr, shifting its COM position and velocity by up to $\sim$100~kpc and a few tens of km\,s$^{-1}$ \citep{Patel2025}. In our four-body model, the first M33–M31 pericentric passage occurs at a median time of $0.8_{-0.1}^{+0.2}$~Gyr from now, with a pericentric distance of $147.4_{-38.8}^{+38.2}$~kpc. By that time, M31 is expected to respond with a median COM displacement of $26.4_{-8.9}^{+11.0}$~kpc and a velocity offset of $42.7_{-17.6}^{+11.0}$~km\,s$^{-1}$.
For the eventual MW–M31 merger, however, the more relevant quantity is the perturbation to the MW–M31 relative orbit rather than the absolute response of M31 alone. In this relative sense, the effect becomes smaller, corresponding to a median MW–M31 displacement of only $10.7_{-5.1}^{+9.3}$~kpc and a relative velocity offset of $30.1_{-12.9}^{+22.1}$~km\,s$^{-1}$. After this pericentric passage, 21\% of the realizations result in M33 merging with M31 before the MW–M31 merger, with a median time of $4.4_{-0.8}^{+1.2}$~Gyr, whereas 5\% merge only afterward ($7.6_{-1.7}^{+1.6}$~Gyr). In most realizations (74\%), M33 does not merge with M31 within the integration time. This outcome likely reflects the relatively large transverse velocity of M33 with respect to M31 implied by the fiducial priors ($\sim170$~km\,s$^{-1}$).

\begin{figure*}[htbp!]
  \centering
    \includegraphics[width=\linewidth]{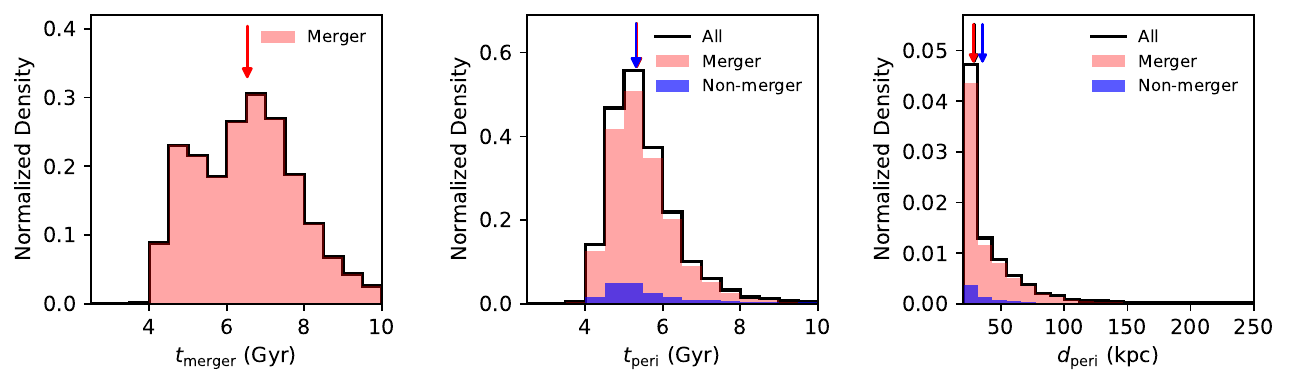}
  \caption{Distribution of merger and pericenter outcomes in the fiducial four-body simulations. Left: normalized probability density of merger times ($t_{\rm merge}$) for merger outcomes among all 10,000 realizations. Middle: normalized probability density of the first pericenter times ($t_{\rm peri}$) for both merging and non-merging outcomes, with the black outline showing all cases. Right: same as the middle panel, but for the distribution of the first pericenter distances ($d_{\rm peri}$). In each panel, the colored arrows indicate the median values of the corresponding distributions. In the middle panel, the three median arrows are nearly overlap due to similar first pericenter times across outcomes.}
  \label{fig:m31_peri}
\end{figure*}

Based on our fiducial four-body model, we further analyse the possible outcomes of the MW–M31 system. For realizations that result in a merger, the distribution of merger times is shown in the left panel of Fig.~\ref{fig:m31_peri}, with a median value of $6.5_{-1.5}^{+1.3}$ Gyr. The middle and right panels show the distributions of the time to the first MW–M31 pericentric passage and the corresponding pericentric separation, evaluated for both merger and non-merger cases. The median first pericentric time is $5.3_{-0.6}^{+0.9}$ Gyr for mergers and $5.3_{-0.6}^{+1.5}$ Gyr for non-mergers. The corresponding median pericentric distances are $28.1_{-8.1}^{+31.5}$ kpc and $35.4_{-12.3}^{+100.4}$ kpc, respectively. Combining all realizations, the first pericentric passage is expected at $5.3_{-0.6}^{+0.9}$ Gyr from now with a distance of $28.9_{-8.9}^{+33.8}$ kpc.

To validate the semi-analytic results obtained under simplified assumptions, we also perform N-body simulations with the Gadget-4 code \citep{Springel2021} for the full four-body configuration. The initial conditions of the galaxies' masses, concentrations, positions and velocities are set to the central values of the fiducial priors listed in Table~\ref{tab:prior}, and the remaining numerical settings followed S25. As shown in Fig.~\ref{fig:Nbody}, the N-body predictions for the MW–M31 merger are broadly consistent with the semi-analytic results. The small remaining differences likely stem from the simplifying assumptions inherent to the semi-analytic treatment.

\subsection{Impact of M31 and M33 Proper Motions on the Predicted Local Group Evolution}\label{subsec:PM_sensi}

\begin{figure}[htbp!]
  \centering
  \begin{minipage}{0.475\textwidth}
    \centering
    \includegraphics[width=\linewidth]{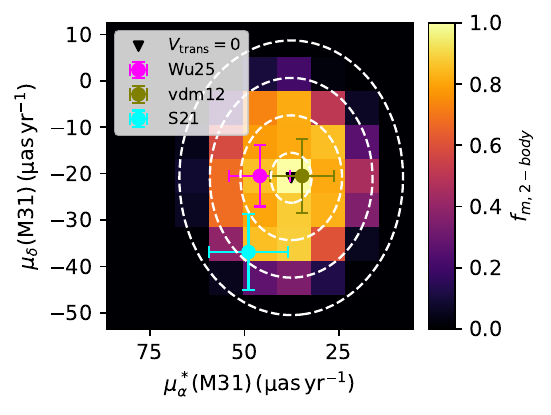}
  \end{minipage}\hfill
  \begin{minipage}{0.475\textwidth}
    \centering
    \includegraphics[width=\linewidth]{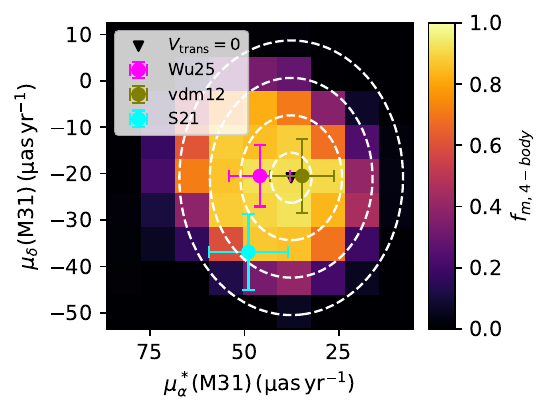}
  \end{minipage}\hfill
  \caption{Sensitivity of the predicted MW–M31 merger probability to the adopted M31 PM for the two-body configuration (MW and M31; top panel) and the four-body configuration (MW, M31, M33 and LMC; bottom panel), shown in the heliocentric frame. Colors indicate the merger probability ($f_{m}$) in each PM grid. The black triangle marks the reference PM of zero transverse velocity relative to the MW. White dashed curves indicate contours of constant M31 transverse velocity (20, 50, 80 and 110 km\,s$^{-1}$ from inner to outer). The outermost contour marks where the transverse and radial components are approximately equal. The fiducial PM adopted in this work \citep{Wu2025} is shown as the magenta symbol, while the HST \citep{vdm2012a} and {\it Gaia} EDR3 \citep{Salomon2021} measurements are indicated in olive and cyan, respectively.}
  \label{fig:pm_sensi}
\end{figure}

The analysis above, together with the comparison to \citetalias{Sawala2025}, shows that both the predicted MW-M31 merger probability and the roles of the satellites depend sensitively on the adopted PMs of M31 and M33. This sensitivity arises from two main effects.
The first is a direct kinematic effect driven by M31’s PM, specifically, the balance between the transverse and radial components of M31’s motion relative to the MW. Smaller transverse velocities naturally yield more radial orbits and thus higher merger probabilities. The second effect arises from the satellites’ influence, where their orbital motions induce reflex motion in the MW and M31 that can either promote or suppress the merger probability.

We examine these two effects separately and quantify their impact on the predicted MW–M31 merger rate. For the first effect, concerning the balance between the radial and transverse components of M31’s motion relative to the MW, we adopt a two-body configuration (MW–M31). The prior on M31 PM is replaced by a uniform distribution spanning $\pm5\sigma$, while all other observables retain their fiducial Gaussian priors truncated at $2\sigma$. This broad prior encompasses all current precise M31 measurements \citep{vdm2012a, Salomon2021, Wu2025} and enlarges the parameter space for testing. As in Section~\ref{subsec:fiducial}, we generate 10,000 initial conditions through MC sampling of these priors and evolve each realization using the methodology described in Section~\ref{sec:method}.

The resulting MW–M31 merger probabilities across the M31 PM grids are shown in the top panel of Fig.~\ref{fig:pm_sensi}, with PMs given in the heliocentric frame. The black triangle marks the point corresponding to zero transverse velocity of M31 relative to the MW, and four contours of constant transverse velocities are overlaid for reference. The merger probability depends sensitively on the adopted M31 PM, ranging from very low values when the transverse velocity is comparable to the radial component to nearly unity as the transverse velocity approaches zero. Consistent with this trend, the three reference measurements \citep{Salomon2021, Wu2025, vdm2012a} trace a progression from $\sim$50\% to nearly 100\% merger probability as the inferred transverse motion decreases, in agreement with \citet{vdm2012b} and \citetalias{Sawala2025}.

We next extend the analysis to the full four-body configuration (MW, M31, M33, and LMC) with the same prior settings and derive the corresponding MW–M31 merger probabilities across the M31 PM grids. The results, shown in the bottom panel of Fig.~\ref{fig:pm_sensi}, exhibit a similar overall trend to the two-body case, but the dependence on M31 PM becomes more complex: a smaller transverse velocity does not always yield a higher merger probability. For example, the \citet{vdm2012a} value, despite implying a smaller transverse velocity than \citet{Wu2025}, produces a lower merger probability. This inversion reflects the influence of satellite-induced reflex motion, whose strength depends on the adopted PMs of both the satellites and M31.

For the influence of satellite-induced reflex motion, we first assess the dependence on the satellites’ own PMs. Since the LMC's PM is accurately measured \citep[e.g.,][]{Kallivayalil2013} and its trajectory is well constrained given fixed values of the other observables, we focus on the potential impact of M33’s PM. Variations in M33’s motion can alter the satellite-induced reflex motion of M31 and thus affect the predicted MW–M31 merger probability. Using the full four-body model and adopting a uniform prior spanning $\pm5\sigma$ for M33 PM, we find that the merger probability is only weakly sensitive to this parameter (Fig.~\ref{fig:m33_pm_sensi}), with nearly all realizations predicting a high merger probability across the explored range. Hence, with all other observables drawn from relatively tight fiducial Gaussian priors, the MW–M31 merger probability depends only weakly on the satellites’ PMs.

\begin{figure}[htbp!]
  \centering
  \begin{minipage}{0.475\textwidth}
    \centering
    \includegraphics[width=\linewidth]{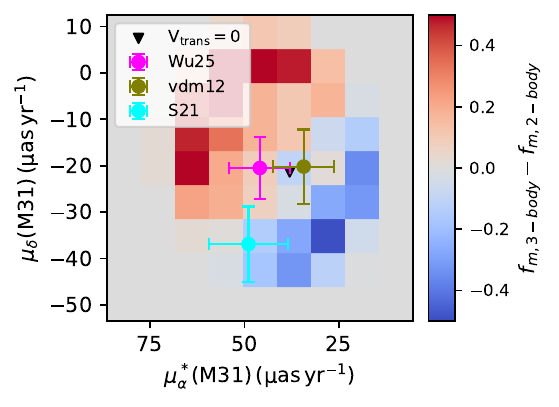}
  \end{minipage}\hfill
  \begin{minipage}{0.475\textwidth}
    \centering
    \includegraphics[width=\linewidth]{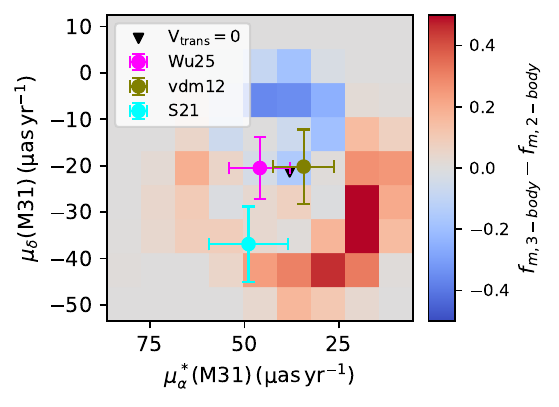}
  \end{minipage}\hfill
  \caption{Dependence of the satellite-induced reflex motion effect on the adopted M31 PM. The color scale indicates the magnitude of such effect, quantified as the difference in merger probability ($f_m$) between the three-body and two-body configurations (three-body minus two-body). In the top panel the additional satellite is the LMC, and in the bottom panel it is M33. Symbol conventions follow those of Fig.~\ref{fig:pm_sensi}.}
  \label{fig:satellite_effect}
\end{figure}

We then examine how the reflex motion effect itself depends on M31’s PM. Keeping all other observables fixed to their fiducial priors, we replace the prior on M31 PM with a uniform distribution spanning $\pm5\sigma$. The reflex motion effect is quantified as the difference in the MW–M31 merger probability between the three-body configurations (MW, M31, and M33/LMC) and the two-body case (MW and M31), as shown in Fig.~\ref{fig:satellite_effect}. Across the explored M31 PM range, this difference varies dramatically, shifting from strong suppression (up to $\Delta f_{m} = -50\%$) to strong enhancement (up to $\Delta f_m = +50\%$). The origin of this dependence is geometric: taking the LMC as an example, the adopted M31 PM sets the orientation of the MW–M31 orbital plane, which determines how the LMC’s phase-space vector projects onto that plane and onto the MW–M31 relative motion, thereby modulating how the LMC’s infall influences the MW–M31 orbit.
Consequently, the LMC-induced reflex motion of the MW can either promote or suppress the MW–M31 merger, depending on the adopted M31 PM. In our fiducial model (Section~\ref{subsec:fiducial}), this reflex motion is oriented primarily along the radial direction within the MW–M31 orbital plane, promoting the merger probability. By contrast, in the scenario explored by \citetalias{Sawala2025}, it projects mostly perpendicular to the orbital plane, increasing the tangential velocity component of the MW-M31 relative motion and thereby suppressing mergers. Fig.~\ref{fig:appendix_track} illustrates this contrast by comparing orbital tracks under these two M31 PM choices for both two-body (MW–M31) and three-body (adding the LMC) configurations.

This geometric dependence also clarifies why, in our sensitivity test using the full four-body model (bottom panel of Fig.~\ref{fig:pm_sensi}), adopting M31 PM values from \citet{vdm2012a} does not reproduce the ``inevitable merger” outcome of \citet{vdm2012b}: in addition to different MW and M31 mass priors, \citet{vdm2012b} did not include the LMC. Under that PM, the LMC-induced barycentric reflex motion of the MW slightly suppresses the merger probability.

For M33, such geometric effect is similar. In Fig.~\ref{fig:appendix_track_m33} we present the orbital tracks under two PM values (\citealt{Wu2025} and \citealt{vdm2012a}), comparing M33's effect under different PM choices.

Overall, although the updated M31 PM measurement from \citet{Wu2025} implies a high MW–M31 merger probability, the fate of the LG, and the future of its two dominant spirals, remains highly uncertain. The dominant source of uncertainty lies in the adopted M31 PM, which affects the outcome both directly and indirectly. Small variations within the currently allowed PM range can lead to large changes in the inferred merger probability. Using the \citet{Wu2025} PM as reference, the merger rate ranges from 73.3\% to 100\% across its $1\sigma$ region, from 64.7\% to 100\% across $2\sigma$, and from 21.4\% to 100\% across $3\sigma$, with all other observational priors held fixed at their fiducial distributions. As an illustrative goal, limiting the inferred merger probability variation to within 10\% across the $2\sigma$ PM range would require reducing the PM uncertainties to $\lesssim 2~\upmu\mathrm{as\,yr^{-1}}$ per component, about a factor of 3–4 smaller than the current typical errors.
Given the strong dependence of the satellite-induced reflex motion on the adopted M31 PM, we emphasize that reliable predictions of the LG fate must be based on at least a full four-body model.
We also note that the semi-analytic model used in this work neglects tidal stripping. As shown in Appendix~\ref{sec:s3}, for the fiducial realization, the corresponding N-body simulation, which includes tidal stripping, yields a very similar MW-M31 merger timescale. This agreement reflects the fact that the merger timescale is primarily governed by the interaction between the two massive haloes, for which stripping of the satellites plays only a minor role. 
However, the treatment of tidal stripping and satellite structure remains simplified even in the N-body simulations, and further improvements will be required for a more realistic description. We therefore expect our main conclusions on the probability of MW-M31 merger to be robust, while satellite-level predictions remain subject to additional uncertainties.

Besides the PMs, the halo masses of the MW and M31 are also key parameters in predicting the future evolution of the LG. Given the range of recent mass estimates for the MW and M31, as reviewed by \citet{Wang2020} and \citet{Bhattacharya2025}, we adopt broadened uniform priors of $6.0\times10^{11}\,M_{\odot} \leq M_{200,\rm MW} \leq 2.0\times10^{12}\,M_{\odot}$ and $5.0\times10^{11}\,M_{\odot} \leq M_{200,\rm M31} \leq 3.0\times10^{12}\,M_{\odot}$, respectively. All other parameters retain their fiducial Gaussian priors truncated at $2\sigma$. 
We then generate 10,000 initial conditions for the four-body configuration (MW, M31, M33, and LMC) through MC sampling and evolve them using the methodology described in Section~\ref{sec:method}. The results are shown in Fig.~\ref{fig:mass_sensi}. 
Overall, the MW–M31 merger probability remains high across the explored mass range, generally exceeding $\sim$80\%. This indicates that the merger probability is not strongly sensitive to the MW/M31 halo masses under the updated M31 and M33 PMs.

\section{Summary}\label{sec:summary}
In this work, we reassess the future dynamical evolution of the LG using updated high-precision PMs for M31 and M33 within the semi-analytic framework of \citetalias{Sawala2025}. The initial conditions are generated by MC sampling of 20 observables from Gaussian priors truncated at 2$\sigma$, and evolved in a four-body system comprising the MW, M31, the LMC, and M33. In the fiducial model, the MW–M31 merger fraction within 10 Gyr is 90\%, with a median merger time of $6.5_{-1.5}^{+1.3}$ Gyr, restoring the previously accepted picture that M31 is very likely to merge with the MW in the future \citep{vdm2012b}. The LMC inevitably merges with the MW before the first MW–M31 encounter, with a median time of $1.3_{-0.4}^{+0.6}$ Gyr, which renders the MW–M31 orbit more radial and increases the merger probability. By contrast, M33 is less likely to merge with M31 (21\% before and 5\% after the MW–M31 merger), and thus has only a minor influence on the overall MW–M31 merger rate.

Furthermore, we find that the predicted MW–M31 merger rate is highly sensitive to the adopted M31 PM values, driven by two mechanisms. First, M31’s transverse velocity directly determines the balance between radial and tangential motion in the MW–M31 orbit: a smaller transverse component naturally leads to a more radial trajectory and a higher merger probability. Second, a satellite-mediated geometric effect arises because the M31 PM sets the orientation of the MW–M31 orbital plane, thereby modulating how the satellite-induced barycentric reflex motions project onto that plane, which can either promote or suppress the merger likelihood. 
Given this strong sensitivity, current PM precision, with typical uncertainties of \(\sim8\,\upmu\mathrm{as\,yr^{-1}}\) per component, is insufficient to draw a firm conclusion about the ultimate fate of the LG. Using the \citet{Wu2025} values as reference, the MW–M31 merger probability still spans 64.7\%–100\% across the $2\sigma$ region. To reach a decisive conclusion, for instance, if the variation in the inferred merger probability across the 2$\sigma$ PM range is kept below 10 percentage points, the required uncertainties would be \(\lesssim 2\,\upmu\mathrm{as\,yr^{-1}}\) per component.
Forthcoming data releases such as {\it Gaia} DR4, the continued long-baseline HST observations, and upcoming wide-field surveys of nearby galaxies with the Chinese Space Station Telescope (CSST), are expected to reduce the PM uncertainties toward these targets and enable a more decisive assessment of the MW–M31 merger scenario.

\section*{acknowledgments}
This work acknowledges the supports from National Key R\&D Programme of China (Grant No. 2023YFA1608303 and 2024YFA1611903) and the National Science Foundation of China (NSFC Grant No. 12422303, 12090040 and 12090044) and the Fundamental Research Funds for the Central Universities (Grant No. 118900M122, E5EQ3301X2 and E4EQ3301X2).

\appendix

\section{Comparison between the N-body simulations and the semi-analytic integration model}\label{sec:s3}
\renewcommand{\thefigure}{A\arabic{figure}}
\setcounter{figure}{0}

\begin{figure*}[htbp!]
  \centering
  \includegraphics[width=\textwidth]{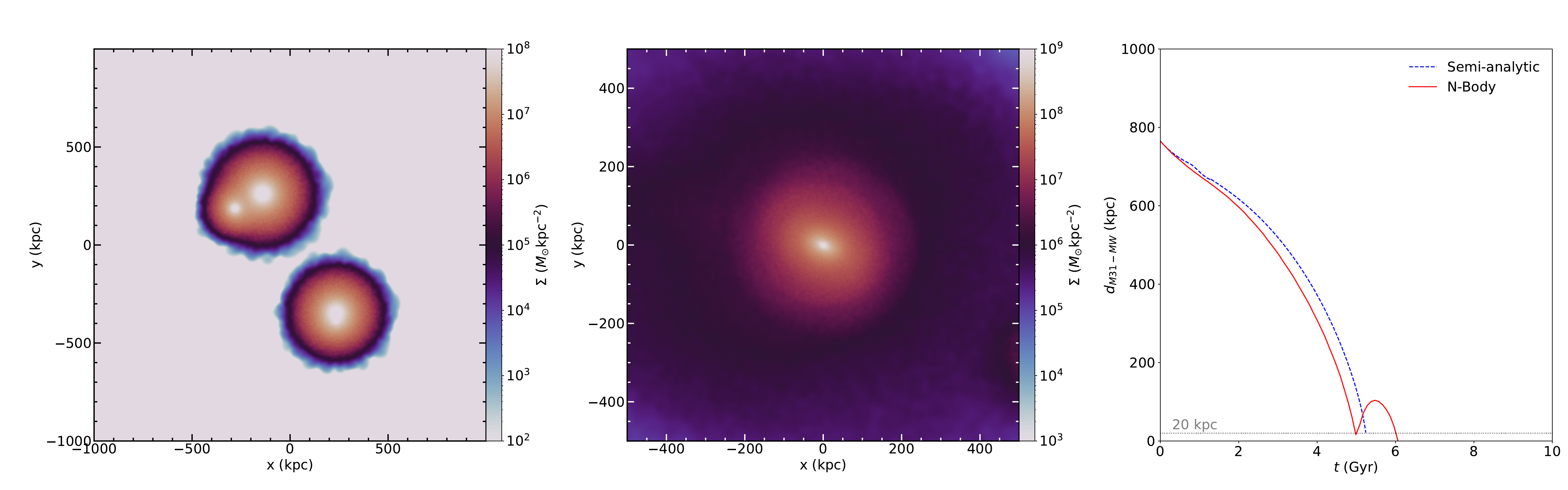}
  \caption{Comparison between the N-body simulations and the semi-analytic model in the four-body configuration. The initial conditions for the galaxies’ masses, concentrations, positions, and velocities are set to the central values of the fiducial priors listed in Table~\ref{tab:prior}.  
  Left panel: projected mass density at the present-day snapshot ($t=0$) in the centre-of-mass (COM) frame for N-body simulations. The MW lies at the bottom right, with the LMC near the MW’s center; M31 and M33 are in the top left.
  Middle panel: projected mass density at the 10 Gyr in the future in the COM frame, by which time the MW has merged with M31. 
  Right panel: MW–M31 distance as a function of time from the N-body simulations (solid red line) and the semi-analytic integration (dashed blue line). Overall, the two approaches yield consistent merger outcome.}
  \label{fig:Nbody}
\end{figure*}

\section{Impact of M33 Proper Motions on the Predicted Local Group Evolution}
\renewcommand{\thefigure}{A\arabic{figure}}
\setcounter{figure}{1}

\begin{figure*}[htbp!]
    \centering
    \includegraphics[width=0.45\linewidth]{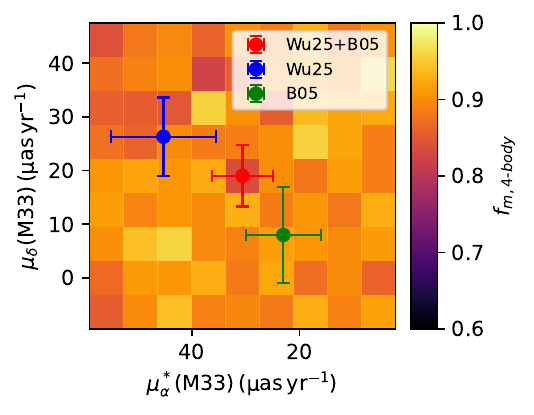}
  \caption{Sensitivity of the predicted MW-M31 merger rate to the choice of M33 PM, shown in the heliocentric frame. Colors indicate the merger probability in each PM grid. The adopted PM in this work is shown as the red symbol, with the \citet{Wu2025} and \citet{Brunthaler2005} measurements indicated in blue and green, respectively.}
  \label{fig:m33_pm_sensi}
\end{figure*}

\newpage
\section{LMC-Induced Reflex Motion under Different M31 Proper Motions}\label{sub:appendix_lmc}

Here we present an example of how LMC-induced reflex motion can have opposite effects on the MW–M31 merger under two representative choices of the M31 PMs, as shown in Fig.~\ref{fig:appendix_track}. Changing the M31 PM rotates the MW–M31 orbital plane and alters the projection of the LMC’s phase-space vector, which yields different MW–LMC interaction geometries with respect to the MW-M31 orbit. In the \citet{Wu2025}–like case (top panels) the LMC-induced reflex motion promotes the radial component and promotes a merger. In the \citet{Salomon2021}–like case (bottom panels) it increases the out-of-plane component and suppresses the merger.
\renewcommand{\thefigure}{A\arabic{figure}}
\setcounter{figure}{2}
\begin{figure*}[htbp!]
  \centering
    \includegraphics[scale=0.75]{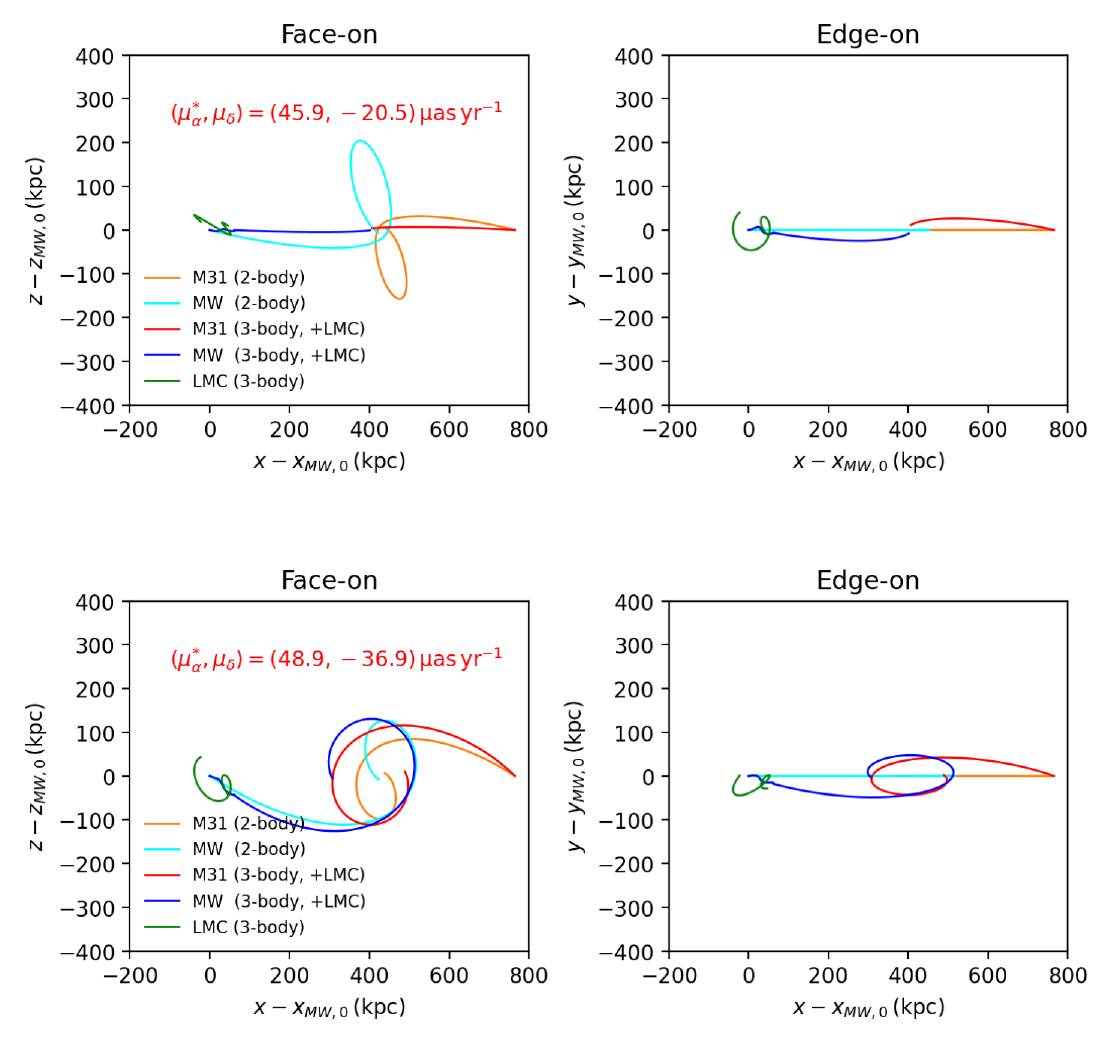}
    \caption{Effect of the LMC-induced reflex motion on the MW-M31 orbital tracks for two representative M31 PMs (top: values from \citealt{Wu2025}, used in our fiducial model; bottom: values from \citealt{Salomon2021}, used in \citetalias{Sawala2025}'s fiducial model). Tracks are projected face-on (left) and edge-on (right) relative to the MW-M31 orbital plane, which defined by the initial positions and velocities of the MW and M31. Curves compare the two-body configuration (MW and M31 only) with the corresponding three-body case including the LMC. In the \citet{Wu2025}-like cases, the LMC-induced reflex motion makes the trajectory more radial and promotes the MW-M31 merger, whereas in the \citet{Salomon2021}-like case, it increases the tangential velocity component and suppresses the merger.}
  \label{fig:appendix_track}
\end{figure*}

\newpage
\section{M33-Induced Reflex Motion under Different M31 Proper Motions}
Similar to Appendix~\ref{sub:appendix_lmc}, here we present an example of how M33-induced reflex motion can have different effects on the MW–M31 merger under two representative values of the M31 PM, the \citet{Wu2025} value (top panels) and the \citet{vdm2012a} value (bottom panels), as shown in Fig.~\ref{fig:appendix_track_m33}. In the latter case, the M33-induced reflex motion introduces a stronger non-radial component into the MW–M31 relative motion.

\renewcommand{\thefigure}{A\arabic{figure}}
\setcounter{figure}{3}

\begin{figure*}[htbp!]
  \centering
    \includegraphics[scale=0.75]{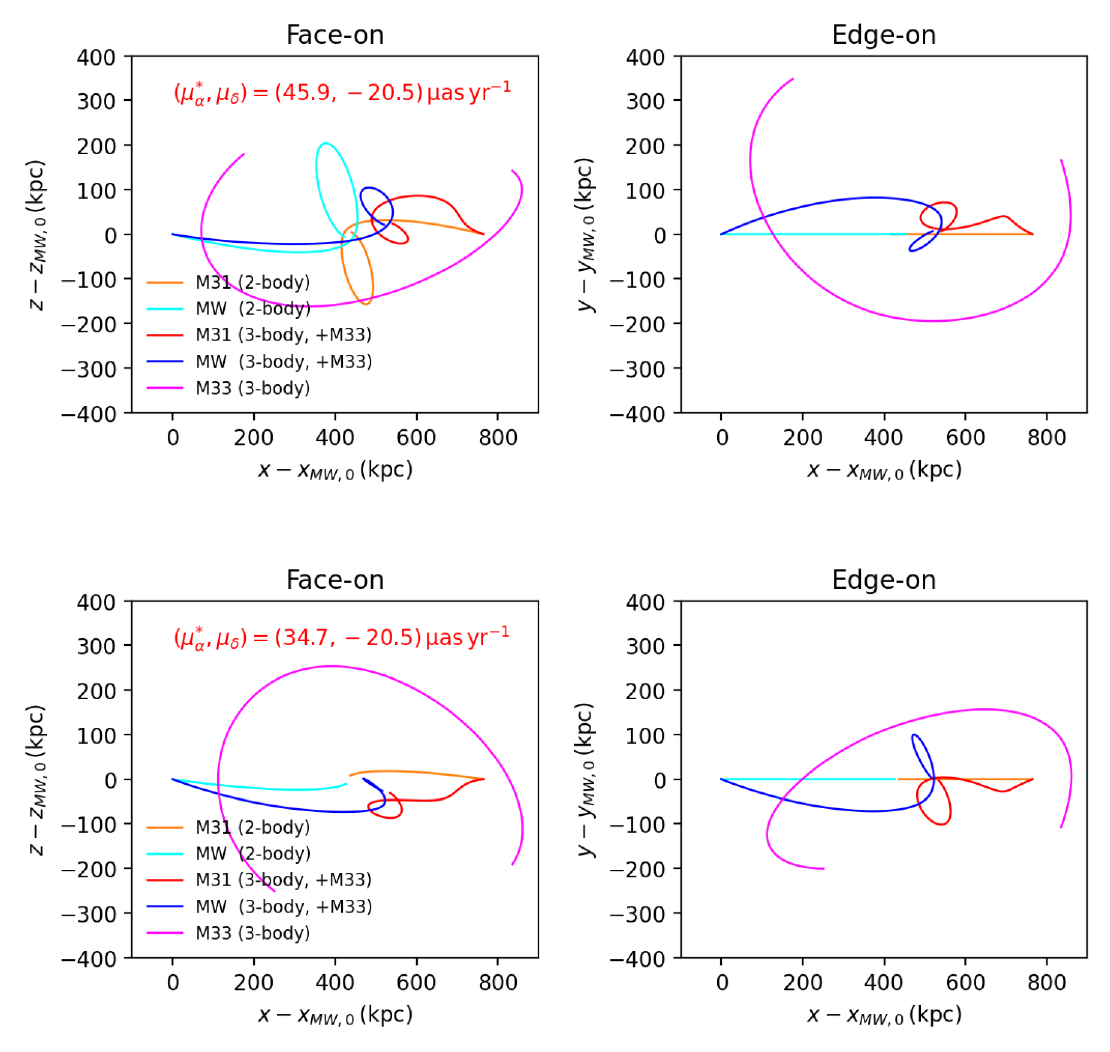}
    \caption{Effect of the M33-induced reflex motion on the MW-M31 orbital tracks for two representative M31 PMs (top: values from \citealt{Wu2025}, used in our fiducial model; bottom: values from \citealt{vdm2012a}, used in \citealt{vdm2012b}'s fiducial model). Tracks are projected face-on (left) and edge-on (right) relative to the MW-M31 orbital plane, which defined by the initial positions and velocities of the MW and M31. Curves compare the two-body configuration (MW and M31 only) with the corresponding three-body case including the M33. In the \citet{vdm2012a}-like cases, the M33-induced reflex motion makes the trajectory less radial.}
  \label{fig:appendix_track_m33}
\end{figure*}

\newpage
\section{Impact of MW and M31 Masses on the Predicted Local Group Evolution}
\renewcommand{\thefigure}{A\arabic{figure}}
\setcounter{figure}{4}
\begin{figure*}[htbp!]
    \centering
    \includegraphics[width=0.45\linewidth]{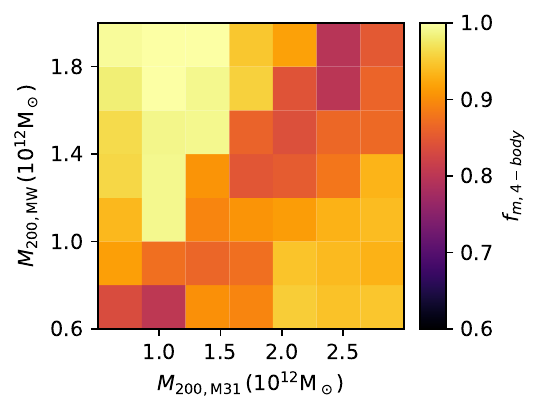}
  \caption{Sensitivity of the predicted MW–M31 merger probability to the halo masses of the MW and M31 in the four-body configuration (MW, M31, M33, and LMC). Colors indicate the merger probability, $f_{\rm m}$, at each point in the mass grid.}
  \label{fig:mass_sensi}
\end{figure*}

\newpage
\bibliography{ref}
\bibliographystyle{aasjournal}

\end{document}